\documentclass[twocolumn,superscriptaddress,prl]{revtex4}
\usepackage{amsmath,amssymb,graphicx,braket,dsfont}
\DeclareMathOperator{\Tr}{Tr}
\usepackage[latin1]{inputenc}
\usepackage[caption=false]{subfig}

\begin{document}
\title{Replica Symmetry Breaking in Cold Atoms and Spin Glasses}

\author{P. Rotondo}
\affiliation{Dipartimento di Fisica, Universit\`a degli Studi di Milano and INFN, via Celoria 16, 20133 Milano, Italy}
\author{E. Tesio}
\affiliation{SUPA and Department of Physics, University of Strathclyde, 104 Rottenrow East, G04NG Glasgow, UK}
\altaffiliation[]{Currently at Orc Group, London, UK}
\author{S. Caracciolo}
\affiliation{Dipartimento di Fisica, Universit\`a degli Studi di Milano and INFN, via Celoria 16, 20133 Milano, Italy}

\begin{abstract}
We consider a system composed by $N$ atoms trapped within a multimode cavity, whose theoretical description is captured by a disordered multimode Dicke model. We show that in the resonant, zero field limit the system exactly realizes the Sherrington-Kirkpatrick model. Upon a redefinition of the temperature, the same dynamics is realized in the dispersive, strong field limit. This regime also gives access to spin-glass observables which can be used to detect Replica Symmetry Breaking. 
\end{abstract}
\maketitle

\section{Introduction}
Replica Symmetry Breaking (RSB) appeared for the first time as a necessary ingredient to solve the Sherrington-Kirkpatrick (SK) model for spin glasses~\cite{Sherrington:PRL:75}, an Ising model characterized by a fully connected network and quenched random interactions. 
This model was introduced to be exactly solvable and not to reproduce a physical system.
Nonetheless, through the years we have accumulated a number of examples of complex problems in biology, informatics, and economy in which RSB is found to play a fundamental r\^{o}le~\cite{Mezard:Spinglass:86}. \\
\indent One of the reasons why the SK model received particular attention is that it allows for a solution via the celebrated Parisi \emph{Ansatz}~\cite{Parisi:PRL:79}. In a nutshell, Parisi suggested RSB as a consistent scheme to break the permutational symmetry of fictitious copies of the system (introduced with the \textit{replica trick}). Physically, RSB in disordered spin systems is interpreted with the emergence of a spin-glass phase characterized by many pure states organized in an ultrametric structure~\cite{Parisi:PRL:83,Mezard:PRL:84}.\\
\indent A fascinating proposal to observe glassy behaviour in a physical system came from the study of light propagation in Kerr-like disordered media \cite{Angelani:PRL:06,Angelani:PRB:06,Leuzzi:PRL:09}, where the slowing-down as the critical point is approached is expected to occur on a much faster timescale than ordinary matter. Progress in this direction is encouraging: for instance, the observation of the mode-locking transition in Random Lasers has been recently reported~\cite{Conti:NatPhot:11}. A scheme to measure the Edwards-Anderson order parameter in interacting-replicas has been presented in~\cite{Morrison:NJP:08} for a Bose gas. Despite these efforts, however, no conclusive results regarding the nature of the spin-glass phase have been presented so far. \\
\indent In the last years, cold and ultracold atoms emerged as a powerful tool to test fundamental models of Condensed Matter physics~\cite{Bloch:NatPhys:05} and disordered systems~\cite{Palencia:NPhys:10,Fallani:PRL:07,Billy:Nat:08, Niederberger:PRL:08}.
Notable attention has been devoted to the Dicke model \cite{Dicke:PhysRev:54}, describing the interaction between $M$ electromagnetic modes and $N$ two-level systems. The superradiant quantum phase transition (QPT) of the single-mode Dicke model was predicted~\cite{Nagy:PRL:10} and observed~\cite{Baumann:Nat:10} in a Bose-Einstein Condensate with cavity-mediated long-range interactions. The appearance of quantum chaos at the Dicke QPT threshold was investigated in~\cite{Emary:PRL:03}, and the Jaynes-Cummings-Hubbard model introduced in~\cite{Greentree:NatPhys:06} can be rewritten as a multimodal Dicke model. This has been recently suggested as a quantum emulator for the fractional quantum Hall effect~\cite{Hayward:PRL:12}.\\
\indent In the spirit outlined above we consider the multimode Dicke Hamiltonian introduced in~\cite{Goldbart:PRL:11,Strack:PRL:11}, where a spin-glass dynamics is obtained for a system of atoms placed in a multimode cavity. In this paper we focus our attention on the possible emergence of RSB in this setup, and the corresponding spin-glass observables. A simple and insightful result is obtained in the resonant, zero field regime (using the terminology adopted in \cite{Goldbart:PRL:11}), where the system exactly realizes the SK Hamiltonian. In the case of a non-zero coupling one can also access the momenta of the overlap distribution and the ultrametric properties which characterize the Replica Symmetric broken phase.   This opens up new interesting opportunities for the validation of spin-glass mean field theories and the observation of spin-glass transitions in a highly controllable system. We also wish to stress here that, from a theoretical standpoint, in the strong-field limit our mapping allows for an exact solution of the multimode Dicke model with quenched disordered interactions.

\section{Model and zero-field limit}

\begin{figure}
\centering%
\includegraphics[scale=0.35]{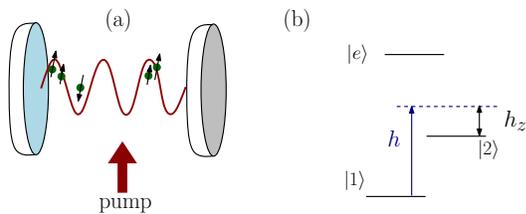}
\caption{ (a) Sketch of the multimode cavity setup. As in Ref.~\cite{Goldbart:PRL:11}, $N$ atoms are placed within a multimode cavity, kept at fixed positions by trapping beams (not shown in the figure) and pumped transversely. Ordering is strongest at the antinodes of the intra-cavity field (red full line), and atoms occupying even antinodes interact ferromagnetically with atoms at even antinodes, and antiferromagnetically with atoms at odd antinodes. (b) Upon adiabatic elimination of the upper state $\Ket{e}$~\cite{Goldbart:PRL:11}, a Dicke interaction is realized by the $\Ket{1}-\Ket{2}$ transition and a field $h$ (Rabi frequency $h_x$, detuning $h_z$). }
\label{fig:setup}
\end{figure}

\begin{figure}
\centering%
\includegraphics[scale=0.35]{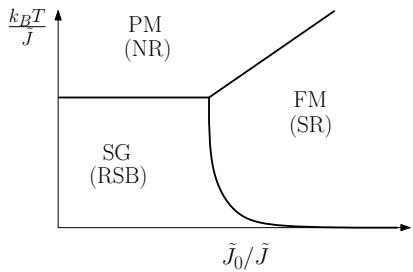}
\caption{Phase diagram for the disordered multimode Dicke model, see~\cite{deAlmeida:JPhysA:78}. At weak disorder (large $\tilde J_0/\tilde J$), a critical temperature is found below which the system is ferromagnetic (FM) and exhibits superradiance (SR). Above this critical temperature and for weak disorder, the system is paramagnetic (PM) and exhibits normal radiance (NR). At relatively low temperatures and strong disorder (small $\tilde J_0/\tilde J$), the system enters the spin-glass (SG) phase and displays RSB.}
\label{fig:phaseDiag}
\end{figure}

The Hamiltonian of the system is a multimode Dicke model with spatially-varying couplings for $M$ photonic modes and $N$ two-level systems~\cite{Goldbart:PRL:11,Strack:PRL:11}:
\begin{align}
H = H_\mathrm{at}+\sum_{m=1}^M \omega_m a^{\dagger}_m a_m +\Omega \sum_{i=1}^N \sum_{m=1}^M g_{im} (a^{\dagger}_m + a_m) \sigma_i^x\,.
\label{eq:MDM}
\end{align}
Here $H_\mathrm{at} = h_x \sum_{i=1}^N \sigma_i^x+ h_z \sum_{i=1}^N \sigma_i^z$, where $h_x$  is the Rabi frequency and $h_z$ is the detuning of the $h$ field, see Fig.~\ref{fig:setup}. The coupling coefficients appearing in the Hamiltonian~(\ref{eq:MDM}) can be finely tuned, offering a high level of control. Disorder is introduced by the presence of many cavity modes, described by the the spatially-varying couplings $g_{im}$. We focus our analysis here on the case where a large number of modes can be supported by the cavity, as in confocal or concentric geometries~\cite{Siegman}.\\
Following \cite{Goldbart:PRL:11,Strack:PRL:11} we proceed by integrating out the photonic modes in order to obtain an effective spin model. In the resonant limit $h_z=0$ (zero field limit), the partition function $Z(N,\beta) = \Tr e^{-\beta H}$ ($\beta$ being the inverse temperature) can be calculated as follows. First we operate a spin-dependent translation to the creation operators (analogous transformations apply to the annihilators):
\begin{equation*}
a^{\dagger}_m \rightarrow a^{\dagger}_m + \frac{\Omega}{\omega_m}\sum_{i=1}^N g_{im} \sigma_i^x \quad \forall\, m = 1, \dots, M\,.
\end{equation*}
We note that these transformations leave unaltered the commutation relations among the photonic modes. Using these new variables the partition function can be put in the form $Z(N,\beta) = Z_{FB}(N,\beta)\, Z_{SK}(N,\beta)$, where $Z_{FB}$ is a free boson partition function and $Z_{SK}$ is given by: 
\begin{align}
&Z_{SK}(N,\beta) = \sum_{\sigma_1 = \pm 1} \cdots \sum_{\sigma_N = \pm 1} e^{-\beta \mathcal{H}_{SK}}\,,\nonumber \\
&\mathcal{H}_{SK} = -\sum_{i,j=1}^N J_{ij}\sigma_i \sigma_j + h_x \sum_{i=1}^N \sigma_i\,
\label{eq:H_SK}
\end{align}
where the $M$-dependence is encoded in the local couplings:
\begin{equation}
J_{ij} \left(M,\{\omega_m\}\right) = \Omega^2 \sum_{m=1}^M \frac{g_{im} g_{jm}}{\omega_m}\,.
\label{eq:J_ij} 
\end{equation} 
The Hamiltonian~(\ref{eq:H_SK}) describes an Ising model with spatially varying couplings in an external magnetic field. When $M \rightarrow \infty$, by the central limit theorem \cite{Strack:PRL:11} the $J_{ij}$'s  become independent random gaussian variables, and are distributed according to: 
\begin{equation*}
P\left(J_{ij}\right) = \frac{1}{(2\pi)^{1/2} J} \exp\left[(J_{ij}-J_0)^2/2J^2\right]\,.
\end{equation*}
We note that in order to obtain relevant disorder fluctuations in the thermodynamic limit ($N \rightarrow \infty$), we must require that $J_0 = \tilde{J}_0/N$, $J = \tilde{J}/\sqrt{N}$, $\tilde{J}_0$ and $\tilde{J}$ being intensive quantities. $\tilde J_0$ and $\tilde J$ parametrizes the disorder introduced by the $g_{im}$, their ratio representing a control parameter for the system (see Fig.~\ref{fig:phaseDiag}). We remark that this condition implicitly imposes large number of modes ($M\sim N$) for the observation of spin-glass transitions, see also~\cite{Goldbart:PRL:11}. Since the couplings $g_{im}$ evolve on the timescale of atomic motion, while the relevant light-atoms interactions occur on a much faster timescale, the random $g_{im}$ coefficients are frozen in a single realization of the system. As a consequence, $\mathcal H_{SK}$ is exactly the Hamiltonian of the Sherrington-Kirkpatrick model~\cite{Sherrington:PRL:75} with an external field $h_x$ (which does not play a fundamental r\^{o}le in what follows). We therefore conclude that in the resonant regime the thermodynamic properties of the disordered Dicke model~(\ref{eq:MDM}) are described by the partition function $Z_{SK}$, so that the system~(\ref{eq:MDM}) effectively realizes the SK model. The phase diagram for this model is well-known~\cite{deAlmeida:JPhysA:78} and displays a spin-glass phase, so that RSB is expected also for the disordered multimode Dicke model~(\ref{eq:MDM}) in the resonant regime (see Fig.~\ref{fig:phaseDiag}).\\
\newline
We now wish to turn our attention to the case of non-zero $h_z$. Restricting to a single photonic mode ($M = 1$) with uniform couplings ($\Omega g_i = g\,,\, \forall\, i$), the resonant case reduces to the fully connected Ising model and displays a classical paramagnetic (PM) to ferromagnetic (FM) phase transition. The only effect of introducing a non-zero external field $h_z$ is the appearance of a threshold in the interaction strength $g^2>h_z$ for the occurrence of the PM/FM transition~\cite{Wang:PRA:73, Lieb:AnnPhys:73,Lieb:PRA:73}. Since the atomic density enters the expression of $g$, this suggests that in our disordered multimode case a non-zero $h_z$ might introduce a threshold for the atomic density below which the phase is always paramagnetic, but this is not expected to change in a qualitative way the existence of a spin-glass phase. Indeed, as discussed below the system still realizes the SK model in the dispersive regime, with $h_z$ acting as a relevant quantity in the detection of RSB.

\section{Dispersive regime and RSB}

To gain a first qualitative insight into the dispersive regime we consider the partition function for non-zero $h_z$ and we use the Golden-Thompson inequality: 
\begin{equation}
\Tr[e^{-\beta(X+Y)}] \leq \Tr[e^{-\beta X} e^{-\beta Y}]\,,
\label{eq:GT}
\end{equation}
which is valid for Hermitian operators $X$ and $Y$. Assuming the inequality to be saturated in Eq.~(\ref{eq:GT}) and splitting the original Hamiltonian~(\ref{eq:MDM}) as $X = h_z\sum_i \sigma_i^z$, $Y = H - X$, we recover the same bosonic decoupling as in the resonant limit and the partition function for the effective spin model can be approximated as:
\begin{equation}
Z(N,\beta) \simeq Z_{FB} \Tr \left[e^{\beta \sum_{ij}J_{ij}\sigma^x_i \sigma^x_j} e^{-\beta h_z \sum_i \sigma^z_i}\right]\,.
\label{eq:Zapprox}
\end{equation}
In the following we will neglect the $h_x$ term for simplicity, but our results are easily extended to the $h_x\neq 0$ case, leaving our conclusions unaffected. The symbol ``$\Tr$" has to be intended as the trace over the $2^N$ dimensional Hilbert space of the spins, the photonic modes being already integrated out. We remark that Eq.~(\ref{eq:GT}) is saturated by  requiring an appropriate relation between $h_z$ and $\Omega$, namely $\beta\Omega^2 = \lambda \tanh{(2\beta h_z)}$. This is a standard result in the context of the Hamiltonian formulation of spin models, such as the classical Ising model~\cite{Mussardo:Statphys:10}. Given the partition function in the form~(\ref{eq:Zapprox}), we are now able to establish a close connection with the usual observables employed in the characterization of the spin-glass phase. The key point in understanding this correspondence consists in rewriting the spin-glass observables in a transfer matrix language. Following \cite{Parisi:PRL:83}, at fixed disorder it is possible to introduce an overlap between pure states (thermodynamic phases) $\alpha$, $\beta$~\cite{Ruelle:StatMech:69}:
\begin{equation*}
q_{\alpha \beta} = \frac{1}{N} \sum_{i=1}^N m_i^{\alpha} m_i^{\beta}\,, \quad m_i^{\alpha} = \braket{\sigma_i}_{\alpha}\,,
\end{equation*}
where the thermal average $\braket{\cdot}_{\alpha}$ has to be intended only on configurations belonging to the pure state $\alpha$. Given the number $S$ of pure states of the system and $P_{\alpha}$ the probability that a typical configuration belongs to the state $\alpha$, the probability distribution for two configurations to have an overlap $q$ is given by:
\begin{equation}
P(q) = \sum_{\alpha, \beta = 1}^S P_{\alpha} P_{\beta}\, \delta\left(q - q_{\alpha \beta}\right)
\label{eq:Pq}
\end{equation}
and acts as an order parameter for the spin-glass transition~\cite{Parisi:PRL:83}. Intuitively, $q_{\alpha\beta}$ measures the `similarity' between the thermodynamic phases $\alpha$ and $\beta$. The breaking of the permutational symmetry of the fictitious copies introduced by the replica trick is physically interpreted as the proliferation of pure states with different macroscopic properties and different overlaps. Hence, in the spin-glass phase $P(q)$ has a non-trivial behavior if Replica Symmetry is broken. In particular, the distribution $P(q)$ can be proven to be equivalent to the probability distribution of the overlap between fictitious replicas~\cite{Parisi:PRL:83}, which can be probed when computing the SK dynamics. We remark that $P(q)$ has been proven to be accessible in Monte Carlo simulations~\cite{Billoire:JPhysA:00,Hukushima:JPSJ:96, Palassini:PRL:99}, and has the property of being a non self-averaging quantity in the presence of RSB~\cite{Mezard:Spinglass:86}. The momenta $\braket{q^n} = \int dq\, q^n P(q)$ of the overlap distribution $P(q)$ can be calculated in a very physical way, introducing two replicated Hamiltonians of the SK model which interact ferromagnetically:
\begin{equation*}
\mathcal H_{2} = \mathcal H_{SK}\left[\sigma^{(1)}\right] + \mathcal H_{SK}\left[\sigma^{(2)}\right] - 2 y \sum_{i=1}^N \sigma_i^{(1)} \sigma_i^{(2)}\,.
\end{equation*} 
The corresponding partition function $Z_2$ can in fact be shown to be a generating function for the momenta $\braket{q^n}\sim\left[\partial \log Z_2/\partial y^n\right]_{y=0}$~\cite{Parisi:PRL:83}.
\begin{figure}
\centering%
\includegraphics[scale=0.35]{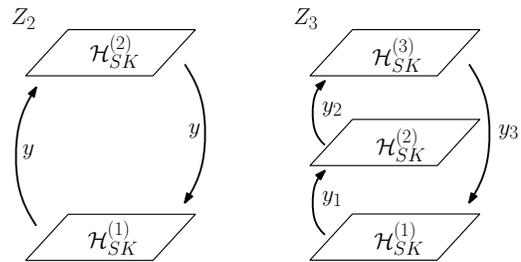}
\caption{Graphical representation of the interacting-replicated partition functions, with $Z_2$ on the left and $Z_3$ on the right. Each layer represents a $\mathcal H_{SK}$ copy, interacting ferromagnetically with another replica with coupling strength $y_i$.}
\label{fig3:replicas}
\end{figure}
Another interesting feature of the spin-glass phase, the ultrametric topology of pure states \cite{Mezard:PRL:84}, can be extracted looking at the partition function built with the following three-replicas Hamiltonian:
\begin{align*}
\mathcal H_3 &= \mathcal H_{SK}\left[\sigma^{(1)}\right] + \mathcal H_{SK}\left[\sigma^{(2)}\right] + \mathcal H_{SK}\left[\sigma^{(3)}\right] +\\ 
&-  \sum_{i=1}^N \left(y_1 \sigma_i^{(1)} \sigma_i^{(2)} + y_2 \sigma_i^{(2)} \sigma_i^{(3)} + y_3 \sigma_i^{(3)} \sigma_i^{(1)}\right)\,.
\end{align*}    
The replicated partition functions $Z_2(y)$ and $Z_3(y_1,y_2,y_3)$ can be rewritten within the transfer matrix formalism as $Z_2(y) = \Tr\left[T(y)^2\right]$, $Z_3(y_1,y_2,y_3) = \Tr\left[T(y_1)T(y_2)T(y_3)\right]$, where 
\begin{equation}
T(h) = e^{\beta \sum_{ij} J_{ij} \sigma^x_i \sigma^x_j} e^{\beta h^{\ast} \sum_i \sigma^z_i}
\label{eq:T}
\end{equation}
is the transfer matrix and $h^{\ast}$ is the solution of the equation: $\tanh h^{\ast} = e^{-2\beta h}$.~\cite{Schultz:RevModPhys:64} Graphically, the replicated partition functions can be visualized as different layers interacting with each other through the ferromagnetic coupling $y_i$ as in Fig.~\ref{fig3:replicas}. Since the multimode Dicke partition function~(\ref{eq:Zapprox}) is written as $Z=\Tr[T(h_z^\star)]$ we find that the same operatorial content captures both the disordered Dicke model~(\ref{eq:MDM}) and the interacting-replica systems $\mathcal H_2$ and $\mathcal H_3$. Therefore, a non-zero (generic) $h_z$ enters the definition of the transfer matrix $T$, whose eigenvalues can be used to calculate the momenta of the overlap distribution and gain access to the observables of the spin-glass phase, at least in a Montecarlo simulation. From an experimental point of view the measure of the overlap distribution at fixed disorder proved to be challenging, because it requires in principle the capability to produce at least two copies of the system with the same disorder. A proposal in this direction came, for instance, in the context of Ref.~\cite{Morrison:NJP:08} for Bose glasses. Essentially, the main idea we wish to convey is that the multimode Dicke model realizes SK in the resonant limit ($h_z=0$), but switching on an additional field allows one to obtain information on the RSB phase via the overlap distribution $P(q)$, without having to create interacting copies of the system. We remark in fact that in our approach there are not two replicated SK hamiltonians interacting with each other as in the original Parisi works, but rather a single hamiltonian with an additional parameter ($h_z$) playing the r\^{o}le of the coupling $y$. It would be nice to find at least one experimental observable in the unreplicated system which allows to gain information about the overlap distribution.\\
\newline
\indent The previous discussion relies on the approximation taken in the Golden-Thompson inequality~(\ref{eq:GT}), and is therefore valid for intermediate values of $h_z$. We now wish to take into examination the dispersive limit $h_z\gg \Omega$, where as in the resonant case $h_z=0$ we will find that the disordered multimode Dicke model realizes a SK dynamics.\\ 
\indent Let us consider the original partition function $Z(N, \beta)$ for non-zero $h_z$ and insert an identity in the form $\mathds{1} = e^{\beta X} e^{-\beta X}$, where $X=h_z\sum _i\sigma_i^z$ as above. Applying the Baker-Campbell-Haussdorf formula (BCH), in the limit $h_z \gg \Omega$ the only contributions come from commutators in the form: 
\begin{equation*}
[\beta X,[\beta X,[\cdots [\beta X,[\beta X,H]]\cdots]\,. 
\end{equation*}
By making use of the explicit form of $H$ we see that at first order $[X,H] \propto \sum_{ik} g_{ik} (a_k+a^{\dagger}_k)\sigma^y_i$, while $[X,[X,H]] \propto \sum_{ik} g_{ik} (a_k+a^{\dagger}_k)\sigma^x_i$, thus showing that these terms can be exactly resummed leading to the partition function
\begin{equation}
Z_\mathrm{disp}(N,\beta) = \Tr[e^{-\beta \tilde{H}} e^{-\beta h_z \sum_i \sigma^z_i}]\,,
\label{eq:Zdisp}
\end{equation} 
where the effective Hamiltonian $\tilde{H}$ is given by:
\begin{equation*}
\tilde{H} =H_0 + \Omega \sum_{i,m} g_{im}(a_m + a^{\dagger}_m)(A(\beta h_z)\sigma_i^x + B(\beta h_z) \sigma_i^y)\,.
\end{equation*}
Here we defined $H_0 = \sum_{m=1}^M \omega_m a^{\dagger}_m a_m $, while $A$ and $B$ are two functions whose Taylor series is determined through the explicit BCH calculation. An appropriate rotation of the Pauli matrices can be performed to recover the original form of the interaction $\sim \Omega_\mathrm{EFF}(a_m^\dagger + a_m)\sigma_i^x$, provided that the coupling strength is rescaled as $\Omega_\mathrm{EFF}(\beta h_z) = \Omega \sqrt{A^2 +B^2}$. Factorizing again the free boson partition function $Z_{FB}$ as above, we find that $Z_\mathrm{disp}$ exactly reduces to the partition function~(\ref{eq:Zapprox}). Alternatively, one can absorb the coupling $\Omega$ into the temperature as $\beta\to\bar\beta=\beta\Omega_\mathrm{EFF}^2$. Given the partition function~(\ref{eq:Zdisp}) we now make use again of the transfer matrix formalism and write it as $Z_\mathrm{disp}=\Tr T(h_z^\star)$. The transfer matrix is in the form $T=V_2 V_1$, and its elements can be explicitly written as~\cite{Mussardo:Statphys:10}
\begin{align*}
\Braket{\sigma_1^{\phantom'}\dots\sigma_N^{\phantom'}|V_1|\sigma_1'\dots\sigma_N'} &= \prod_{k=1}^N e^{-\beta h_z \sigma_k^{\phantom'}\sigma_k'}\\
\Braket{\sigma_1^{\phantom'}\dots \sigma_N^{\phantom'}|V_2|\sigma_1'\dots \sigma_N'} &= \prod_{i=1}^N\delta_{\sigma_i^{\phantom'},\sigma_i'}\prod_{i,j=1}^N e^{\bar\beta J_{ij} \sigma_i^{\phantom'}\sigma_j'}\,.
\end{align*}
With these definitions, the trace operation reduces to a classical sum over the spin configurations $\{\sigma\}$ and we obtain
\begin{widetext}
\begin{align}
Z_\mathrm{disp} &= \Tr\left(V_2 V_1\right) = Z_{FB}\sum_{\{\sigma\}\{\sigma'\}}\Braket{\sigma_1^{\phantom'}\dots \sigma_N^{\phantom'}|V_2|\sigma_1'\dots \sigma_N'}\Braket{\sigma_1'\dots \sigma_N'|V_1|\sigma_1^{\phantom'}\dots \sigma_N^{\phantom'}} = \nonumber\\
&=Z_{FB}\sum_{\{\sigma\}\{\sigma'\}}e^{\bar \beta \sum_{ij}J_{ij}\sigma_i^{\phantom'}\sigma_j^{\phantom'}}e^{-\beta h_z\sum_j \sigma_j^{\phantom'}\sigma_j'}\prod_{i=1}^N \delta_{\sigma_i^{\phantom'}\sigma_i'} = e^{-N\beta h_z}Z_{FB} Z_{SK}(N,\bar\beta)\,.
\end{align}
\end{widetext}
The effective spin model emerging from the disordered multimode Dicke model~(\ref{eq:MDM}) in the dispersive regime is therefore given again by the SK model, upon redefining the temperature as $\beta\to\bar\beta$. Once again, reintroducing $h_x$ does not change this result in a qualitative way. The connection established above with spin-glass observables is therefore confirmed in the dispersive limit, as the partition function is in the form $Z_\mathrm{disp}=\Tr T$.
\\
\indent The derivation presented above shows that in the strong-field (dispersive) regime the SK model is exactly retrieved from a multimode Dicke dynamics. However, we note that in the regime $h_z\gg\Omega $ the spin glass phase is not accessible, because the system is well below the usual strong coupling threshold of the Dicke model. This would in fact result in an effective temperature $\bar\beta$ whose value never approaches the critical one of the SK model. The main point we wish to make here is that the disordered Dicke model is thermodynamically equivalent to SK in both the zero-field ($h_z=0$) and strong-field ($h_z\gg\Omega$) regimes. This suggests that this connection extends also for generic and intermediate values of $h_z$, as discussed earlier in this Section (see the discussion after Eq.~(\ref{eq:T})), in the same way as the multimode Dicke model with the same couplings is equivalent to a ferromagnetic fully-connected Ising model \cite{Lieb:AnnPhys:73,Wang:PRA:73,Lieb:PRA:73}. The results presented here are intended to be the first step in this direction.

\section{Conclusions}

We analyzed a multimode Dicke model with quenched disorder, recently proposed for cold atoms in cavity  setups~\cite{Goldbart:PRL:11,Strack:PRL:11}. Spin-glass dynamics and frustrated interactions are expected, and we are able to prove that in the resonant (zero-field) regime the system exactly realizes the paradigmatic SK model (as already anticipated in the context of Ref.~\cite{Goldbart:PRL:11}). Quite surprisingly, in the dispersive (strong-field) regime this result stays unaffected upon a redefinition of the temperature. Moreover, for non-vanishing values of the coupling the operatorial content of the multimode Dicke model gives access to the spin-glass observables which characterize the Replica Symmetry broken phase. In the strong-field limit the equivalence between the multimode Dicke model and the SK model once again becomes exact, but the spin glass phase is not physically accessible. However, our work suggests that the connection between the SK and the multimode Dicke models extends into the domain of intermediate couplings, which will be the focus of future work.\\
The system offers a high degree of tunability and control, and we stress that the dispersive regime might be more accessible experimentally as absorption and radiation pressure are reduced. From a theoretical standpoint, our approach provides an exact, strong-field solution of the multimode Dicke model with quenched disorder. With a view to the study and validation of spin-glasses mean field theory, dispersive cavity-mediated long range interactions in cold atomic gases appear as a promising benchmark for future research, as they allow for the physical realization of the paradigmatic SK model for spin glasses. The detection (in experiments or in Monte Carlo simulations) of the overlap distribution would in fact give information on the Replica Symmetry Broken phase in a highly tunable and controllable physical system.

Financial support from the Leverhulme Trust (for ET, research grant F/00273/0) is gratefully acknowledged.

\bibliography{PietroBibliography}
\end{document}